# Second waves, social distancing, and the spread of COVID-19 across America

Karl J. Friston[1], Thomas Parr[1], Peter Zeidman[1], Adeel Razi[2], Guillaume Flandin[1], Jean Daunizeau[3], Oliver J. Hulme[4,5], Alexander J. Billig[6], Vladimir Litvak[1], Cathy J. Price[1], Rosalyn J. Moran[7] and Christian Lambert[1]

[1]*The Wellcome Centre for Human Neuroimaging, University College London, UK*
[2]*Turner Institute for Brain and Mental Health & Monash Biomedical Imaging, Monash University, Clayton, Australia*
[3]*Institut du Cerveau et de la Moelle épinière, INSERM UMRS 1127, Paris, France*
[4]*Danish Research Centre for Magnetic Resonance, Centre for Functional and Diagnostic Imaging and Research, Copenhagen University Hospital Hvidovre, Kettegaard Allé 30, Hvidovre, Denmark.*
[5]*London Mathematical Laboratory, 8 Margravine Gardens, Hammersmith, UK*
[6]*Ear Institute, University College London, UK*
[7]*Centre for Neuroimaging Science, Department of Neuroimaging, IoPPN, King's College London, UK*

***E-mails***: *Karl Friston, k.friston@ucl.ac.uk; Thomas Parr, thomas.parr.12@ucl.ac.uk; Peter Zeidman, peter.zeidman@ucl.ac.uk; Adeel Razi, adeel.razi@monash.edu; Guillaume Flandin, g.flandin@ucl.ac.uk; Jean Daunizeau, jean.daunizeau@googlemail.com; Ollie Hulme, oliverh@drcmr.dk; Alexander Billig, a.billig@ucl.ac.uk; Vladimir Litvak, v.litvak@ucl.ac.uk; Rosalyn Moran, rosalyn.moran@kcl.ac.uk; Cathy Price, c.j.price@ucl.ac.uk; Christian Lambert, christian.lambert@ucl.ac.uk*

## Abstract

We recently described a dynamic causal model of a COVID-19 outbreak within a single region. Here, we combine several of these (epidemic) models to create a (pandemic) model of viral spread among regions. Our focus is on a second wave of new cases that may result from loss of immunity—and the exchange of people between regions—and how mortality rates can be ameliorated under different strategic responses. In particular, we consider hard or soft social distancing strategies predicated on national (Federal) or regional (State) estimates of the prevalence of infection in the population. The modelling is demonstrated using timeseries of new cases and deaths from the United States to estimate the parameters of a factorial (compartmental) epidemiological model of each State and, crucially, coupling between States. Using Bayesian model reduction, we identify the effective connectivity between States that best explains the initial phases of the outbreak in the United States. Using the ensuing posterior parameter estimates, we then evaluate the likely outcomes of different policies in terms of mortality, working days lost due to lockdown and demands upon critical care. The provisional results of this modelling suggest that social distancing and loss of immunity are the two key factors that underwrite a return to endemic equilibrium.

## Contents

# Introduction

This technical report describes a dynamic causal model of how COVID-19 spreads among regions by combining multiple models of an epidemic within a single region. This (epidemic) model was introduced recently to showcase the potential of variational Bayesian procedures in fitting mechanistic or generative models of an epidemic to measurable outcomes (Friston et al., 2020).

Mechanistic or generative models enable forecasting based upon epidemiological parameters estimated from empirical data (Wu et al., 2020), e.g., serial interval estimates (Nishiura et al., 2020; Yang et al., 2020). Models of this sort have been applied to the initial outbreak in Wuhan, China (Sun et al., 2020; Wang et al., 2020a; Wang et al., 2020c) to predict new cases or clinical resources (Moghadas et al., 2020) and the effects of various interventions (Prem et al., 2020; Wells et al., 2020). Such models are invaluable when predicting the demand for critical care (Ferguson et al., 2020; Moghadas et al., 2020). The particular contribution of this report is to illustrate the use of variational procedures to compare different models of the same data—and the inherent latitude for building large multifactorial models that can handle different kinds of outcomes.

The (dynamic causal) model (DCM) used in this report has a degree of predictive validity (Friston et al., 2020) and Bayesian model comparison suggests that it provides a simpler and more accurate account than equivalent single factor (e.g., SEIR) models (Moran et al., 2020). However, the DCM only considered a single outbreak in a single region and therefore precluded epidemiological trajectories that feature things like second waves. In this technical report[1], we build upon this *epidemic* model to create a *pandemic* model

---

[1] This technical report is a follow-up to an original report prepared in anticipation of the RAMP (Rapid Assistance in Modelling the Pandemic) initiative (https://royalsociety.org/topics-policy/Health-and-wellbeing/ramp/). It should be read as a proof of concept whose main aim is didactic; namely, to explain how variational procedures enable highly parameterised compartmental models to be inverted and compared quickly and efficiently. As such, it

that comprises multiple epidemic (single region) models. In what follows, we apply this model to explain regional timeseries from the United States, treating each State as a region and modelling the exchange of people—who may or may not be infected—among States. Our focus is on the interplay between regional outbreaks in the evolution of the pandemic and how this evolution informs strategic responses, such as social distancing. Models of the international spread of COVID-19—such as the Global Epidemic and Mobility Model (GLEAM)—usually partition the world into regions centred on major transportation hubs (e.g., airports). These regions are then connected by the flux of people travelling daily among them (Chinazzi et al., 2020). See also (Steven et al., 2020; Wu et al., 2020). In what follows, we apply the same idea to daily travel between the United States of America, equipping each State with its own epidemiology that becomes coupled through reciprocal exchange of their denizens. This creates a loosely coupled (nonlinear) oscillator model, of the sort that is used widely in other settings: e.g., (Jafri et al., 2016; Kaluza and Meyer-Ortmanns, 2010; Ladenbauer and Obermayer, 2019; Lizarazu et al., 2019; Schumacher et al., 2015).

Often, the first wave of a pandemic, so-called "herald waves" (Simonsen et al., 2018), are followed some months later by a second or third waves of infection (see Figure 1) that may, in some instances, be more severe than the first, such as those seen in the influenza pandemics of 1918 (H1H1), 1957 (H2H2), 1968 (H3N2), and 2009 (H1N1). If one commits to the idea that a generative model of measurable outcomes is necessary to properly predict systemic dynamics, then the natural question is: what causes a second wave? One answer is that a regional population is re-exposed to infection by an influx of infected people from another—that may itself have been caused by the first. Clearly, the degree to which an outbreak in another region induces a second wave in the first will depend sensitively on the level of herd immunity inherited from the first wave. It is therefore important to consider the degree to which herd immunity is lost following the first wave; either through an endogenous loss of immunity within the first population or a renewal of that population with people who are not immune. For example, about 0.5% of the American population move between States every day[2]. This movement 'mixes' the total population, with consequent loss of herd immunity.

Heuristically, the picture that emerges can be likened to a Californian or Australian wildfire with embers seeded throughout a large territory. One ember may gain a sufficient hold to cause the first flareup and become the epicentre of a bushfire. As the fire rages through the locale, consuming combustible material, it will eventually burn itself out. However, if the right conditions prevail—and embers are carried to another locale—a second fire will start and thereby elicit a chain of fires. In this analogy, the second wave corresponds to a reignition of the first by embers from the second. However, there is a natural fire retardant in the first locale (i.e., herd immunity) that offers some protection (i.e., there is nothing left to burn).

---

provides the technical details that allow people to reproduce the modelling at home. Alternatively, it could complement epidemiological modelling with Approximate Bayesian Computation. Although the authors of the current report are involved in the clinical management of COVID-19 patients—and are experts in biological timeseries modelling—they are not virologists or epidemiologists. As such, the validity of the model described in this report may or may not be endorsed by experts in the appropriate fields.

[2] According to the Bureau of Transportation Statistics (http://www.transtats.bts.gov/), a total of 631,939,829 passengers boarded domestic flights in the United States in the year 2010. This corresponds to 1.73 million passenger flights per day. The population of the United States is about 327.2 million; a ratio of 190.

Crucially, this protection will decline over time as new growth furnishes more combustible material (i.e., a loss of immunity). So, what causes a second wave—is it the recursive spread of the fire (i.e., pandemic) or is it susceptibility to reignition (i.e., herd immunity)?

These considerations highlight the importance of herd immunity and the factors that underwrite resistance to infection. After the first wave, immunity in any given population may be lost for several reasons. These include a natural loss of immunity of the sort seen in seasonal influenza—mediated by viral mutation or the prevalence of different viral serotypes. For COVID-19, this is an area of active research that has yet to provide definitive answers. However, empirical evidence suggests that neutralising antibodies to COVID-19 can be raised fairly quickly (Bao et al., 2020; Bendavid et al., 2020; Chan et al., 2013). Furthermore, instances of reinfection are sufficiently low to suggest that immunity to COVID-19 may be long lasting, at least over a period of months. Furthermore, studies in nonhuman primates suggest it is difficult to elicit the symptoms of COVID-19 after an initial infection (Bao et al., 2020)[3]. However, this does not necessarily guarantee an enduring herd immunity.

The transmission of SARS-CoV-2 depends on many factors, including seasonal variation in transmission strength, the duration of immunity, and cross-immunity with other coronaviruses. SARS-CoV-2 belongs to the betacoronavirus genus, which includes the SARS, MERS, and two other human coronaviruses, HCoV-OC43 and HCoV-HKU1 (Kissler et al., 2020; Su et al., 2016). HCoV-OC43 and HCoV-HKU1 infections may be asymptomatic or produce mild to moderate upper respiratory tract symptoms; namely, a common cold (Kissler et al., 2020). Accumulating evidence suggests that primary SARS-CoV-2 infection causes a mild illness in the majority of cases with a minority progressing to severe lower respiratory infection, interstitial pneumonia, acute respiratory distress syndrome (ARDS) and multiple organ failure. The primary infections may also be further complicated by additional insults, such as secondary bacterial infections or thromboembolic events[4]. Immunity to HCoV-OC43 and HCoV-HKU1 appears to be lost over a few months. However, betacoronaviruses can induce immune responses against each another. For example, SARS can generate neutralizing antibodies against HCoV-OC43 that can endure for years (Chan et al., 2013), while HCoV-OC43 infection can generate cross-reactive antibodies against SARS (Chan et al., 2013). Now, the question is: does SARS-CoV-2 behave like SARS, conveying long-lasting immunity or does it behave like HCoV-OC43, conferring immunity for just a few months? In what follows, we will consider both scenarios and the implications for how we live with COVID-19—or not.

This report tries to characterise the interplay between population fluxes and the waning of immunity using a compartmental model of ensemble dynamics. Typically, these kinds of models employ a sparse coupling between nonlinear systems (here a compartmental model of a regional outbreak). In the present setting, the sparse coupling is a small exchange of people between States that can be parameterised in terms of the daily probability that 'I will leave one State for another'. Clearly, this will depend upon whether I can leave home, which depends upon any social distancing currently in play.

---

[3] After seroconversion, asymptomatic monkeys were challenged with a second dose of SARS-CoV-2. Neither viral loads in nasopharyngeal and anal swabs nor viral replication in primary tissue compartments was evident in re-exposed monkeys. On the basis of follow-up virologic, radiological and pathological assessment, monkeys with re-exposure showed no recurrence of COVID-19. Bao, L., Deng, W., Gao, H., Xiao, C., Liu, J., Xue, J., Lv, Q., Liu, J., Yu, P., Xu, Y., Qi, F., Qu, Y., Li, F., Xiang, Z., Yu, H., Gong, S., Liu, M., Wang, G., Wang, S., Song, Z., Zhao, W., Han, Y., Zhao, L., Liu, X., Wei, Q., Qin, C., 2020. Reinfection could not occur in SARS-CoV-2 infected rhesus macaques. bioRxiv, 2020.2003.2013.990226.

[4] See https://www.sciencemag.org/news/2020/04/how-does-coronavirus-kill-clinicians-trace-ferocious-rampage-through-body-brain-toes for an accessible summary of COVID-19's reach beyond the lungs.

This brings us to the second focus of the current modelling; namely, how does social distancing mitigate interregional influences? To answer this question, one has to have a formal model of social distancing. There are two attributes of any social distancing—its functional form and its argument, i.e., what is social distancing a function of? In what follows, we consider functional forms based on dual criteria: namely, the prevalence of measurable infection, and demand upon critical care. In brief, these functions can be regarded as *adaptive strategies*. In other words, they are social distancing responses to changes in measurable quantities. These responses may be mediated via governmental advice, social media, behavioural dispositions, and other affordances[5] that we can lump together in terms of a propensity to avoid interpersonal contact—or not.

We will consider two threshold-based responses to the prevalence of infection in the population. This prevalence is fairly straightforward to estimate on the basis of currently available tests for the presence of the virus on mucosa. This does not require exhaustive or comprehensive testing; provided these measures are modelled appropriately (e.g., using a generative model of the sort considered below). Although offering a straightforward model of social distancing there remains an outstanding issue: do I (or the government) base my social distancing on the prevalence of infection in my region, or is it driven by the experience of other regions in my country. In other words, should social distancing be based upon regional or national criteria, and therefore enacted at a regional or national level? In what follows, we fit a pandemic model to regional data from the United States and ask what would happen if a national (i.e., Federal) social distancing strategy was adopted, as opposed to a regional (i.e., State) approach.

This report comprises three sections. The first rehearses the dynamic causal model presented previously, with an emphasis on the extension to multiple regions—and implicit coupling among regions. The second section presents the results of model fitting to timeseries from States in America[6]. This section considers the sensitivity of cumulative deaths to various model parameters, with a focus on the connectivity among States that shapes the overall progression of the pandemic. The final section considers social distancing strategies by simulating what would happen under regional and national approaches, with hard (e.g., lockdown) and soft (e.g., partial) social distancing.

---

[5] In the sense of Gibson, J.J., 1977. The theory of affordances, in: R, S., Bransford, J. (Eds.), Perceiving, acting, and knowing: Toward an ecological psychology. Erlbaum, Hillsdale, NJ, pp. 67-82. For example, seeing the person on the pavement affords the opportunity to circumnavigate them by 6 feet.

[6] Available from https://github.com/CSSEGISandData/COVID-19.

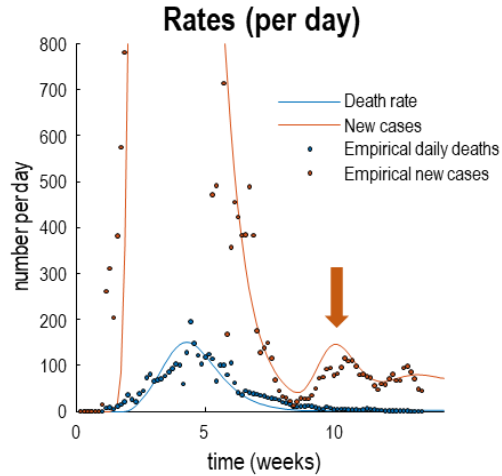

**Figure 1: second waves in China**. This figure illustrates a secondary wave of new cases in reports from China. The dots represent empirical records of new cases and deaths as a function of weeks from the onset of the outbreak. The lines correspond to the predicted incidences, under a (single region) model described in (Friston et al., 2020). These data are presented to illustrate a resurgence of new cases (orange arrow), several weeks after the first wave that is not accompanied by a marked increase in death rates. We will see a similar phenomenology in predictions of new cases and deaths for the United States in subsequent figures.

# Dynamic causal modelling

The dynamic causal model used in this report is an extension of a compartmental model of a regional outbreak detailed in (Friston et al., 2020). This (epidemic) model is a factorial extension of a conventional SEIR (*susceptible*, *exposed*, *infected*, and *recovered*) model in which the four states[7] of the SEIR model are unpacked into four factors (*location*, *infection*, *symptoms*, and *testing*), each with four states. This furnishes a $4^4 = 256$-compartmental model that allows for different combinations of states to generate data. For example, being infected does not necessarily mean that one has to manifest symptoms. Conversely, one can be severely ill without having a viral infection[8]. Similarly, including a testing factor allows for people

---

[7] We will use State to indicate a *State* of the United States and *state* to denote a level of a factor in the generative model.

[8] For example, acute respiratory distress due to secondary bacterial pneumonia, which may itself be antimicrobial resistant: https://www.southcentre.int/wp-content/uploads/2020/03/RP-104.pdf.

to be contagious but not reported as testing positive for SARS-CoV-2[9].

The factorial structure of this model exploits conditional independencies among the factors to provide a relatively straightforward parameterisation. Technically, it uses a mean field approximation to certain dependencies. For example, the probability of developing symptoms depends on, and only on, whether I have an infection. However, the probability of becoming infected does not depend on whether I am symptomatic or not. The four factors in question are shown schematically in Figure 2 in the form of a compartmental model. The accompanying parameters of this compartmental model (that mediate the conditional dependencies) are listed in Table 1. This model is described in detail in (Friston et al., 2020) and used subsequently to look at herd immunity and lockdown cycles in (Moran et al., 2020). Here, the model is equipped with two further features: namely, a loss of immunity over time and a distinction between susceptible and non-susceptible members of the population. We equipped the model with a loss of immunity, with the following parameter:

$$\theta_{imm} = \exp(-1/\tau_{imm}) \tag{1.1}$$

This controls a slow flux from a state of *immunity* to a state of *susceptibility*, under the *infection* factor. We made the provisional assumption that immunity to SARS-CoV-2 resembles the immunity to SARS-CoV-1. Operationally, we modelled this as a period of immunity with a time constant of 32 months: c.f., (Kissler et al., 2020). This parameter was included to repeat simulations under more pessimistic conditions, in which immunity could be lost within a few (four) months (c.f., a common cold human coronavirus, HCoV-OC43).

The distinction between susceptible and non-susceptible was modelled by assigning people to a *susceptible* or *resistant* state, such that they did not participate in the spread of the virus if they were resistant. We assumed *a priori* that half of the total population would be susceptible to infection and estimated the proportion of resistant cases under the model[10]. This can be regarded as a crude approximation to varying levels of susceptibility in the population at large and could be further refined as more is learnt about the susceptibility to COVID-19 infection; for example, by leveraging regional demographic data.

Numerous viral-evasion mechanisms are known. For example, mutations in the 'resistome' cause susceptibility to infection, and other (yet to be identified) mutations cause resistance to infection. (Beutler et al., 2007). Though age and comorbidity contribute substantially to fatality rates, the host factors that influence resistance or susceptibility to infection with pathogenic human coronaviruses (CoVs) are largely unknown but might involve several mechanisms. For example, innate immune responses to CoVs are initiated by recognition of double-stranded RNA and induction of interferon, which turns on gene expression programs that inhibit viral replication (Heer et al., 2020). Furthermore, epidemiological evidence suggests that SARS-CoV-2 infection in children is less frequent and severe than in adults: age-related ACE2 receptor expression might be a relevant host factor (Cristiani et al., 2020). Including a distinction between susceptible and resistant enables the model to explain data from the total population, as opposed to a susceptible population. This is important when considering models of how the virus is spread among regions by movement of people who may or may not be susceptible and therefore capable of infecting—and being infected by—others.

---

[9]  Undocumented infections may be the source of infection for 79% of documented cases Li, R., Pei, S., Chen, B., Song, Y., Zhang, T., Yang, W., Shaman, J., 2020. Substantial undocumented infection facilitates the rapid dissemination of novel coronavirus (SARS-CoV2). Science, eabb3221.

[10]  This parameter determines the initial conditions and specifies the proportion of the population that do not participate in epidemiological transitions. As such, we omit the state of resistance from subsequent figures, for clarity.

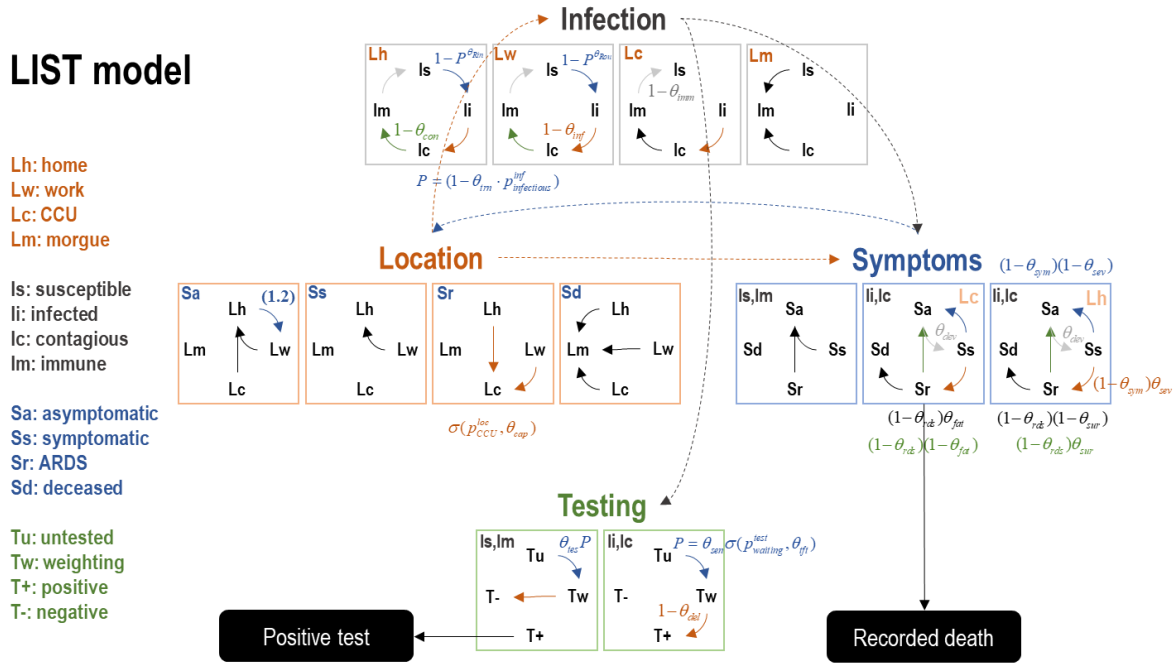

**FIGURE 2**

**Figure 2: the dynamic causal (LIST) model**. This schematic summarises the dynamic causal model that is used to explain timeseries data in a single region. In brief, it comprises four factors, each with four states (listed in the key on the left), giving 256 states or compartments. Regional factors include *location*, *infection*, *symptom*, and *testing* status (the *infection* state of *resistance* has been omitted from the schematic for simplicity). The small arrows denote transitions among states that are parameterised in terms of the probability of moving from one state to another, every day. Black or unlabelled arrows denote a unit probability. Coloured arrows designate transitions that are determined by the model parameters ($\Theta$—please see Table 1). To suppress visual clutter, the expressions for transition probabilities are colour-coded within each factor. The quantities $p^i$ are marginal probabilities over the states of the $i$-th factor. Crucially, transitions among the states of one factor depend upon other factors. These conditional dependencies (highlighted in blue) are illustrated by showing the transition probabilities among the states of one factor under the levels of another. For example, the probability that I will move from a state of being asymptomatic (**Sa**) to being symptomatic (**Ss**) depends upon whether I am infected (**Ii** and **Ic**) or not (**Is** and **Im**). Furthermore, the probability that I will move from ARDS (**Sr**) to being deceased (**Sd**) depends upon whether I am located in a critical care unit (**Lc**) or at home (**Lh**)—and so on. The parameters of the implicit transition probabilities are listed in Table 1 and are described in detail in (Friston et al., 2020). A particularly important transition is the probability that I will leave home (i.e., expose myself to more potentially contagious contacts) on any given day; namely, social distancing. This is denoted by (**1.2**), corresponding to equation (1.3) in the main text.

Figure 3 illustrates how a single-region (epidemic) model can be used to construct a (pandemic) model that encompasses several regions. The key aspect of this factorial extension (i.e., including a region factor) rests upon the coupling among regions. Here, people who are not confined in self-isolation (i.e., at home) are available to travel with a certain daily probability from one state to another. *A priori*, this probability is based upon the probability that any American citizen will fly from one state to another every day. This can be written down as the following transition probability

$$\Delta N = \Delta N_i - \Delta N_k$$
$$\Delta N_k = P(work_t \mid region_k, asymptomatic) \cdot N_k \cdot \theta_{ik}$$
$$\Delta N_i \propto P(work_t \mid region_i, asymptomatic), \quad s.t., \quad \sum \Delta N_i = \sum \Delta N_k \tag{1.2}$$

$$P(work_{t+1} \mid region_i, asymptomatic) = P(work_t \mid region_i, asymptomatic) - \Delta N / N_i$$
$$P(work_{t+1} \mid region_k, asymptomatic) = P(work_t \mid region_k, asymptomatic) + \Delta N / N_k$$

Here, $N_i$ corresponds to the number of people in the *i*-th State and $\Delta N$ corresponds to the relative number of people that are exchanged between States *i* and *k*. This construction ensures that the total number of people in each state does not change over time. In turn, this constraint means that the flux of people between any two States can be parameterised with a single coupling or connectivity parameter, $\theta_{ik}$.

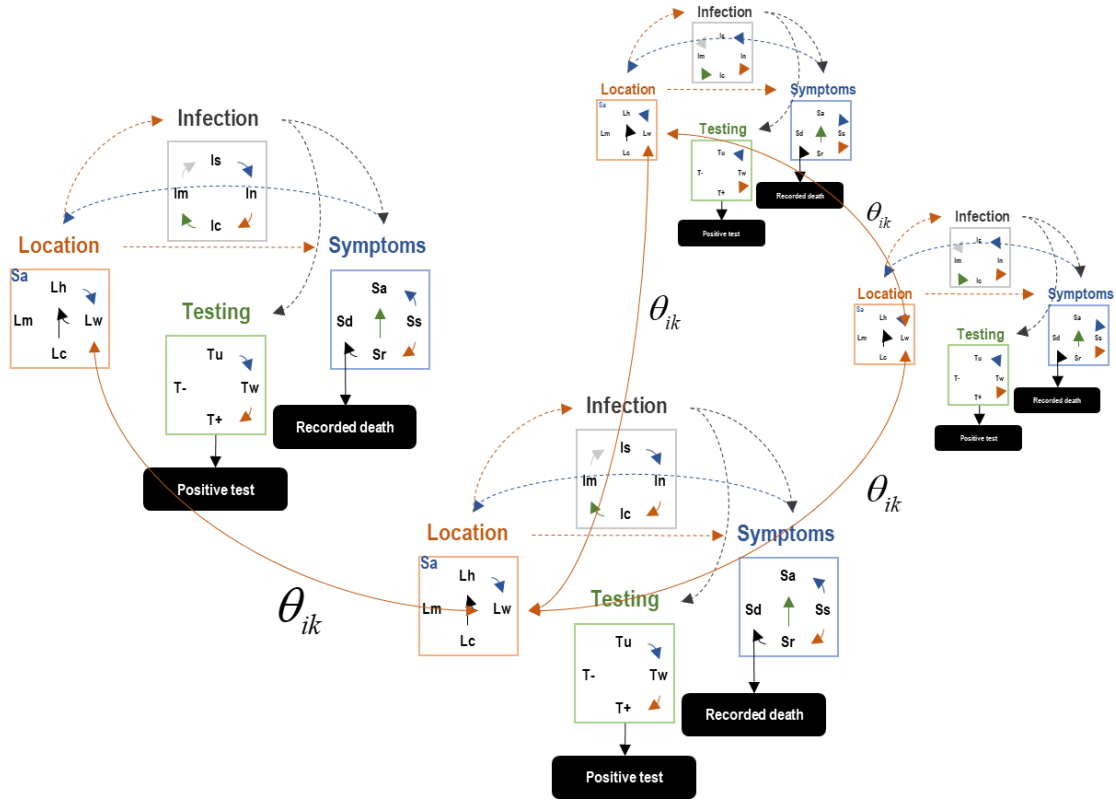

**Figure 3: a multi-region model**. This schematic summarises the dynamic causal model used to explain timeseries data from multiple regions. This (pandemic) model is composed of several regional (epidemic) models. In brief, the model for a single region comprises four factors, each with four states, giving $4^4 = 256$ states or compartments per region. These regional models are then assembled to model the coupling among eight regions giving $256^8$ compartments. However, due to conditional independencies, this can be treated as a collection of 256 compartmental models; providing one links the states of one region to the states of another carefully. Here, this linking or connectivity is parameterised in terms of a probability flux or exchange of people among regional populations. The probability that I will move from one region to another depends upon whether I am at work (i.e., not at home, **Lh**) and do not consider myself to be ill (i.e., I am asymptomatic, **Sa**). In short, the exchange between different regional populations is limited to the people who are not at home and are consequently in a position to travel. This is illustrated by the arrows in the figure that connect the appropriate states. The parameters of interregional coupling correspond to rate constants or effective connectivity that ensure the conservation of total numbers in each region. For example, the probability of moving to New York from New Jersey is the same as a probability of moving from New Jersey to New York; however, the number of people commuting in either direction will depend on the respective population sizes of New Jersey and New York.

Usually, in these kinds of connectivity models, one considers different sparsity constraints on the coupling architecture. One could use Bayesian model comparison based upon the variational free energy (a.k.a. evidence lower bound) to test for different connectivity structures (Friston et al., 2015); e.g., full connectivity, a serial connectivity based upon spatial distance between states, the time of onset and so on. However, for simplicity, we elected to use a model with full connectivity and eliminate redundant connections *post hoc*, using Bayesian model reduction (Friston et al., 2018).

The second important extension to the epidemic model was the inclusion of a dual criteria social distancing process, parameterised as follows:

$$P(work_{t+1} \mid home_t, asymptomatic) = \theta_{out} \cdot \sigma(p_{infected}^{infection}, \theta_{sde}) \cdot \sigma(p_{CCU}^{location}, \theta_{sde} \cdot \theta_{cap} \cdot 8)$$

$$P(work_{t+1} \mid home_t, asymptomatic) = \theta_{out} \cdot \sigma(q_{infected}^{infection}, \theta_{sde}) \cdot \sigma(q_{CCU}^{location}, \theta_{sde} \cdot \theta_{cap} \cdot 8)$$

(1.3)

The first equality parameterises the probability of going to work in the morning as a function of the marginal probabilities of certain states for the region in question, $p$, while the second has the same functional form but takes the marginal probability from all regions, $q$. These can be regarded as a regional (State) and national (Federal) social distancing responses, respectively. It would also be possible to respond based on a linear combination of these two formulations, however we limit the simulations presented here to one or other of the extremes. The states in question are the probability that any member of the population is currently infected or requires critical care. The form of the decreasing threshold (sigmoid) functions that constitute this model of social distancing is illustrated in Figure 4.

Intuitively, this models social distancing as the probability of leaving home, which only attains normal levels when two criteria are met. First, the proportion of the population infected with coronavirus must be below some threshold and second, the number of people in critical care must be less than some fraction of maximum capacity. Equation (1.3) lumps the thresholds together so that there is a single social distancing threshold that is applied to both criteria. When the threshold is low (say 1/32) we effectively have a lockdown policy that will remain in place until less than 1/32 of the population are deemed to be infected and the occupancy of critical care facilities has fallen below 8/32 = ¼ of total capacity. A more liberal threshold could be interpreted as a softer social distancing strategy, where certain (non-vulnerable) people are allowed to return to school (or work) and some degree of social distancing is maintained when commuting or at work (e.g., wearing face masks).

Note that social distancing is modelled as a part of the epidemiological dynamics generating data. In other words, it is not an exogenous input or strategic response that enters the model. Rather, it is installed as a reactive and adaptive process that best explains the observed data. An interesting twist here is that the optimisation of the parameters that shape social distancing are those that best explain the data; namely, the social distancing that has been achieved. They are not optimised to minimise some cost function (e.g., mortality rates)—they are a measure of what a population actually does when confronted with a pandemic. Later, we will simulate different levels of social distancing—by increasing or decreasing the estimated threshold—to see what mitigates morbidity; either in terms of cumulative deaths or economic morbidity (as measured in terms of lost working days under the *location* factor).

Finally, we included a back-to-work parameter that allowed people who were immune to be exempt from social distancing and return to work. This is a fictive aspect of the model, in the sense that it would rest upon knowing whether somebody was immune or not. This enabled us to model the potential benefits of being able to measure seroconversion and allow people back to work if they had detectable (and hopefully neutralising) antibodies to SARS-CoV-2.

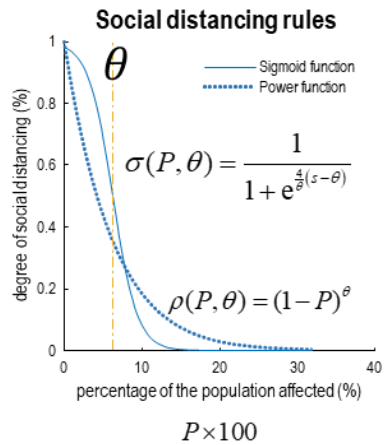

**Figure 4: threshold strategies**. This figure illustrates the different kinds of social distancing functions that could be adopted to model social distancing responses. Both parameterise the degree of social distancing as a function of the proportion of the population that are currently infected (and are currently occupying critical care facilities). This proportion can, in principle, be estimated directly or indirectly, given current testing capabilities. These social distancing functions are decreasing functions of the prevalence of infection (or critical care occupancy). In other words, as prevalence increases, the probability of leaving home (e.g., self-isolation) decreases. The two examples above are distinguished by the form of this decrease. The (threshold) strategy used in this report is based upon a threshold, afforded by the reverse sigmoid function (solid line). Conversely, the (exponent) strategy used in previous models (Friston et al., 2020) decreases smoothly as a power function of prevalence (broken line). The functional forms are given by the equations in the figure. Please see main text for further details.

With this pandemic model in place, we modelled the eight States in America with the greatest number of cases of COVID-19, using standard (variational Laplace) Bayesian methods (Friston et al., 2020). We chose to analyse eight States because the key information in the accompanying timeseries lies in the form of the transients or fluctuations in new cases and deaths. This form is better evinced by larger numbers. Put simply, including States that have yet to experience an epidemic provides no useful information that would inform the parameters that are shared between States.

One advantage of modelling data from the United States of America is that one can, *a priori*, assume that many factors are conserved from State to State, given the homogeneity of the medical infrastructure within America. Unlike the epidemic model, the population size of each state serves as an informative prior on the susceptible population size. In other words, previously we estimated the number of people affected by an epidemic as an unknown quantity. Here, we try to explain the data in terms of all the population who are caught up in the pandemic—an unknown proportion of which may be resistant to infection. Technically, this involves placing very precise shrinkage priors on the population size, corresponding to the population of each State. Finally, because each state is likely to have been seeded with one or two infected individuals at different times, we modelled the pandemic as ensuing from the epicentre; namely, New York State[11]. This involves estimating the number of people in New York who were infected eight days after records began.

To expedite model inversion, we used a further mean field approximation, by assuming conditional independence between the between-region (i.e., connectivity) parameters in Table 1A and the within-region parameters in Table 1B. This allowed us to estimate the parameters for each region (i.e., State) separately and then use the ensuing State-specific estimates to infer the connectivity parameters (and initial case load at the epicentre). This greatly finesses the numerics; however, it comes at the price of ignoring conditional dependencies between the two sorts of parameters. This completes our description of the model. The next section turns to the results of fitting multivariate timeseries of new cases and deaths under this model.

### TABLE 1A

Parameters of the pandemic (multiple region) model, $N(\eta, C)$

(NB: prior means are for scale parameters $\theta = \exp(\vartheta)$)

| Number | Parameter | Mean | Variance | Description |
|--------|-----------|------|----------|-------------|
| 1 | $\theta_n$ | exp(-4) | 1/4 | Number of initial cases in the epicentre |
| 2 | $\theta_{ik}$ | exp(-8) | 1/16 | Population flow between regions |

### TABLE 2B

Parameters of the epidemic (LIST) model and priors, $N(\eta, C)$

(NB: prior means are for scale parameters $\theta = \exp(\vartheta)$)

| Number | Parameter | Mean | Variance | Description |
|--------|-----------|------|----------|-------------|
| 1 | $\theta_n$ | exp(-4) | 1/4 | Number of initial cases |
| 2 | $\theta_r$ | 1/2 | 1/16 | Proportion of resistant cases |
| Location | | | | |
| 3 | $\theta_{out}$ | 1/3 | 1/64 | Prob(*work* \| *home*): probability of going out |
| 4 | $\theta_{sde}$ | 1/32 | 1/128 | Social distancing threshold |
| 5 | $\theta_{cap}$ | 16/100000 | 1/64 | CCU capacity threshold (per capita) |
| Infection | | | | |

---

[11] We considered a series of models with sparse connectivity and full connectivity (entertaining initially infected cases in one State or all States). We chose to report the full connectivity model to showcase the use of subsequent Bayesian model reduction.

| 6 | $\theta_{Rin}$ | 4 | 1/64 | Effective number of contacts: home |
|---|---|---|---|---|
| 7 | $\theta_{Rou}$ | 48 | 1/64 | Effective number of contacts: work |
| 8 | $\theta_{trn}$ | 1/4 | 1/64 | Prob(*contagion* | *contact*) |
| 9 | $\theta_{inf} = \exp(-\frac{1}{\tau_{inf}})$ | $\tau_{inf} = 4$ | 1/64 | Infected (pre-contagious) period (days) |
| 10 | $\theta_{con} = \exp(-\frac{1}{\tau_{con}})$ | $\tau_{con} = 4$ | 1/64 | Contagious period (days) |
| 11 | $\theta_{imm} = \exp(-\frac{1}{\tau_{imm}})$ | $\tau_{imm} = 32$ | 1/64 | Period of immunity (months) |
| **Symptoms** | | | | |
| 12 | $1 - \theta_{dev} = \exp(-\frac{1}{\tau_{inc}})$ | $\tau_{inc} = 8$ | 1/64 | Incubation period (days) |
| 13 | $\theta_{sev}$ | 1/128 | 1/64 | Prob(*ARDS* | *symptomatic*) |
| 14 | $\theta_{sym} = \exp(-\frac{1}{\tau_{sym}})$ | $\tau_{sym} = 5$ | 1/64 | Symptomatic period (days) |
| 15 | $\theta_{rds} = \exp(-\frac{1}{\tau_{rds}})$ | $\tau_{rds} = 12$ | 1/64 | Acute RDS period (days) |
| 16 | $\theta_{fat}$ | 1/2 | 1/64 | Prob(*fatality* | *CCU*) |
| 17 | $\theta_{sur}$ | 1/16 | 1/64 | Prob(*survival* | *home*) |
| **Testing** | | | | |
| 18 | $\theta_{tft}$ | 1/1024 | 1/4 | Threshold: testing capacity (per capita) |
| 19 | $\theta_{sen}$ | 1/1024 | 1/4 | Prob(being tested) (per day) |
| 20 | $\theta_{del} = \exp(-\frac{1}{\tau_{del}})$ | $\tau_{del} = 2$ | 1/4 | Delay in reporting test results (days) |
| 21 | $\theta_{tes}$ | 1/4 | 1/4 | Prob(*tested* | *uninfected*) (per day) |

**Secondary sources** (Huang et al., 2020; Kissler et al., 2020; Mizumoto and Chowell, 2020; Russell et al., 2020; Verity et al., 2020; Wang et al., 2020b) and:

- https://www.statista.com/chart/21105/number-of-critical-care-beds-per-100000-inhabitants/
- https://www.gov.uk/guidance/coronavirus-COVID-19-information-for-the-public
- http://www.imperial.ac.uk/mrc-global-infectious-disease-analysis/COVID-19/

These prior expectations should be read as the effective rates and time constants as they manifest in a real-world setting. For example, a four-day period of contagion is shorter than the period that someone might be infectious (Wölfel et al., 2020)[12], on the (prior) assumption that they will self-isolate, when they realise they could be contagious. Although the scale parameters are implemented as probabilities or rates, they are estimated as log parameters, denoted by $\vartheta = \ln \theta$.

---

[12] Shedding of COVID-19 viral RNA from sputum can outlast the end of symptoms. Seroconversion occurs after 6-12 days but is not necessarily followed by a rapid decline of viral load.

# Results

This section reviews the within and between-State parameters following model inversion under a regional social distancing response. We then use these parameters to predict outcomes that have yet to be observed. In the final section, we will repeat these predictions under different social distancing strategies. Our focus in this section is on how the second wave, if any, is shaped by social distancing responses and the acquisition—and subsequent loss—of immunity.

Figure 5 reports the differences among States in terms of selected parameters for each State, ranging from the population size, through to the probability of testing its denizens. The blue bars report the posterior expectations, while the pink bars are 90% Bayesian credible intervals. Notice that there is no uncertainty about the population sizes because these are known quantities. From the current perspective, the interesting thing to note here is the social distancing threshold that varies among States, around 0.03. In other words, each State is behaving as if lockdown is invoked when either 3% of the population become infected or critical care occupancy approaches $8 \cdot 3\% = 24\%$ of total capacity. Interestingly, California appears to have the most stringent social distancing response so far, with the lowest threshold. This may be relevant later when trying to explain differential mortality rates.

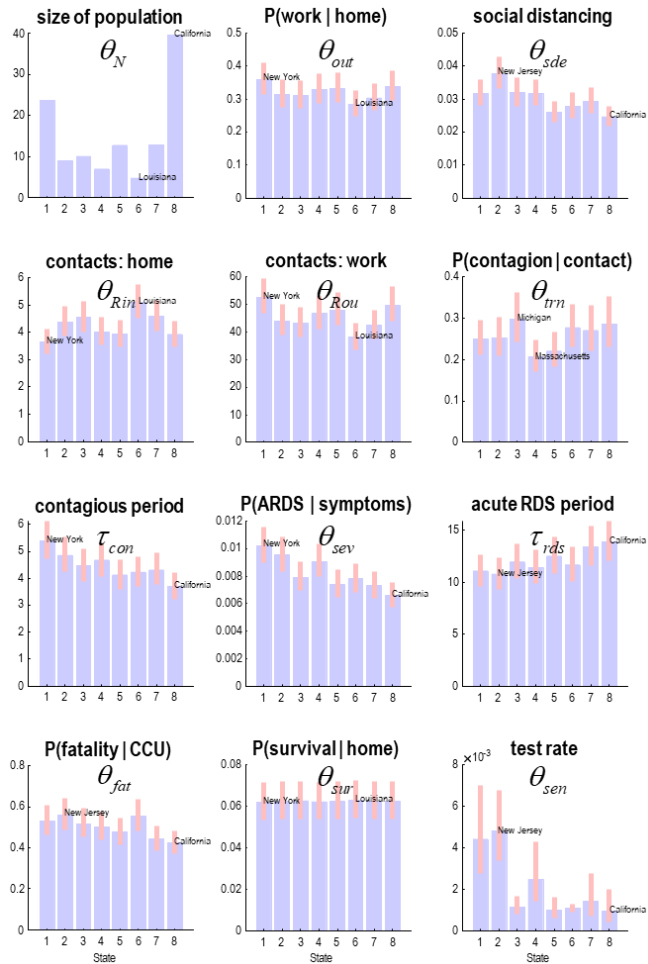

**Figure 5: differences among States**. This figure reports the differences among States in terms of selected parameters of the generative (epidemic) model, ranging from population size, through to the probability of testing. The blue bars represent the posterior expectations, while the pink bars are 90% Bayesian credible intervals. Notice that these intervals are not symmetrical about the mean because we are reporting scale parameters—as opposed to log parameters. For each parameter, the States showing the smallest and largest values are labelled. For example, New Jersey and New York behave as if they had a relaxed social distancing threshold, when compared to California. Please see next figure for a key to the States.

Figure 6 summarises the estimated connectivity or population exchange between States. The upper right panel shows the connectivity among States as an adjacency matrix. The key thing to take from this analysis is that nearly all the connections among states have been removed following Bayesian model reduction, leaving only reciprocal exchange with New York. Clearly, this is not what is happening in the field. However, it is the simplest account of the data at hand. In other words, it is sufficient to describe the current data in terms of one epicentre (New York) exchanging infected individuals with the remaining States. In this analysis, the greatest flux of people appears to be between New York and California. The question now is whether these fluxes have a material impact on cumulative deaths or mortality rates. This sort of question can be addressed using a sensitivity analysis.

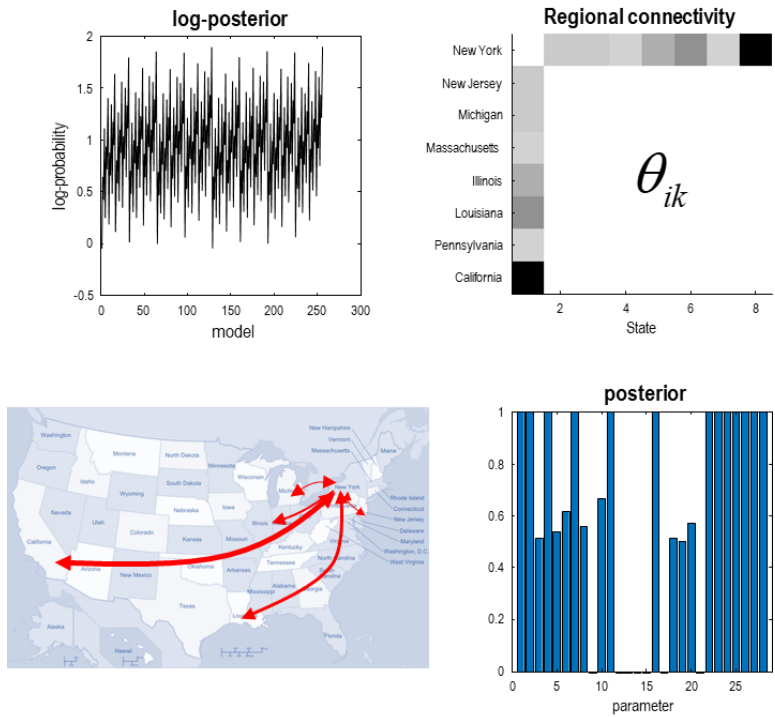

**Figure 6: connectivity and viral spread**. This figure reports the connectivity among states, following Bayesian model reduction of a full connectivity model. The upper left panel shows the log evidence of the 256 models with the greatest

evidence. A model here constitutes a reduced model in which various combinations of connectivity parameters have been removed. The upper right panel shows the maximum a posteriori (MAP) estimates of the ensuing Bayesian model average, as an adjacency matrix. The elements of this matrix quantify the rate or probability that a State in each column will deliver a proportion of its (out-of-home) population to a State in the rows. For example, the greatest flux of people is between New York and California. The lower right panel shows the posterior probability of a model with and without each of these parameters, based upon Bayesian model comparison. For the connectivity parameters involving New York, we can be nearly 100% certain that the model that includes this coupling parameter has greater evidence than a model that does not. The insert on the lower left provides a schematic representation of the connectivity, based upon these estimates. The heavier connectors correspond to a greater probability of moving between States.

A sensitivity analysis involves changing each of the parameters by a small value and measuring the consequent change in cumulative deaths, while holding all the other parameters constant. Figure 7 shows the results of this analysis for the connectivity parameters of the model that survived Bayesian model reduction. Interestingly, the effects of movement between States does not appear to have a consistent effect on cumulative deaths. Sometimes increasing commuter traffic decreases overall mortality and sometimes it is increased. Having said this, the flux that has the greatest (mitigating) effect on overall mortality is the exchange between New York and California. In other words, mixing the populations in New York and California would, under this model, save lives. The explanation for these complicated effects rests upon the intricate dynamics of latent states that constitute the underlying causes of the pandemic.

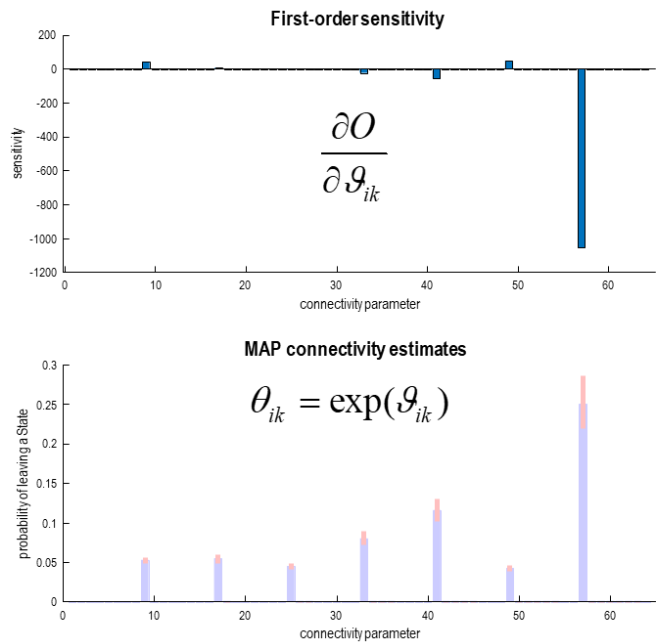

<div align="center">F<span>igure</span> 7</div>

**Figure 7: sensitivity analysis**. This figure reports the effect of changing each connectivity parameter on cumulative deaths over an 18-month period. The upper panel shows the rate of increase (or decrease) in cumulative deaths per unit change in the (log) parameters. These sensitivity metrics are based upon a first

order Taylor expansion about the *maximum a posteriori* values shown in the lower panel. The blue bars correspond to the most likely estimate and the pink bars report the 90% credible intervals. Interestingly, the effects of connectivity or coupling among States are mixed. In some instances, increasing the exchange between one State and another will increase or decrease overall death rates; presumably, based upon the respective capacity of different states to respond to pressure on their clinical care capacities.

Figure 8 provides a prediction of the future course of the pandemic in each of the eight States. The picture that emerges here is what one might expect from a loosely coupled oscillator model (here, a factorial compartmental model of ensemble dynamics). Notable aspects of these predictions are that New York experiences a marked peak in death rates early in the epidemic, whereas other States follow after a week or so. In terms of the epidemiology, an interesting feature of these predictions is a second peak in death rates at around 38 weeks (i.e., 28 weeks after the first peak). Another aspect of these predictions is the large variability in death rates between States. Much of this variability can be accounted for by differences in populations; however, as shown in the lower right panel, mortality rates (as estimated by the expected number of deaths in one year divided by the population of each State) still show a substantial variation. For example, the mortality rate in New York is estimated to be just over 0.1%, while it is much smaller in California. Recall that these predictions are entirely conditioned upon the model and available data. Having said this, the currently available peak death rates in New York and California testify to some difference that cannot be explained in terms of population sizes, e.g., climactic effects on transmission strength that mediate seasonal influences (Kissler et al., 2020). Alternatively, the differential social distancing estimates in New York and California (see Figure 5) may speak to regional or cultural differences (e.g., the prevalence of high density, low-cost housing or ethnic differences in communal activities).

The cumulative deaths under this particular model are consistent with predictions based upon other modelling work. For example, at 20 weeks, the cumulative deaths in the United States has—on some conservative estimates—been reported in the media to be around 60,000[13]. The eight States here constitute 119 million people (about a third of the population). The cumulative deaths within these States are estimated to be about 48,000, under the current model. One potentially reassuring aspect of these results is that predicted (annual) mortality rates in most States is, overall, less than that attributable to seasonal influenza (0.1%) (Paget et al., 2019). Indeed, typical mortality rates appear to be in the order of 0.05% or less. This is in line with projections based upon global data and countries who experienced an early epidemic, such as Italy.

---

[13]  Consistently, across countries, excess death is higher than reported COVID-19 deaths: in many cases deaths in care homes or in the home are not reported systematically as COVID-19 related. For example, https://www.nytimes.com/interactive/2020/04/21/world/coronavirus-missing-deaths.html

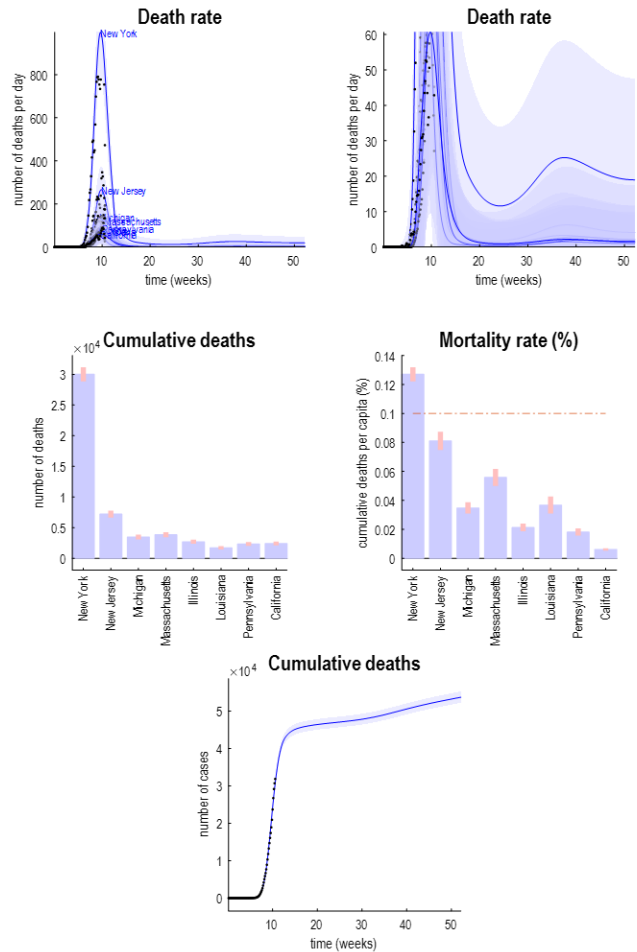

**Figure 8: predicted mortality rates**. This figure reports the predicted trajectory of death rates for each State. The upper panels show the predicted time course of death rates per day under a social distancing strategy based upon regional estimates of prevalence. The expected rates are shown as blue lines, while the shaded blue areas correspond to 90% Bayesian credible intervals. The dots report empirical data observed to date. The same data are shown in the upper right panel after scaling the Y axis. This enables the second wave of deaths—predicted under this model—to be seen more clearly, about 28 weeks following the first wave. The lower left panel shows the cumulative deaths in each State, based upon these predictions, in terms of the total expectation (blue bars) and accompanying 90% credible intervals (pink bars). These projections reflect differences among the States in terms of their population—and their response to the influx and subsequent explosion of infected cases. The middle right panel adjusts for differences in the population by expressing mortality rates (per year) as a percentage of the population of each state. The broken horizontal line corresponds roughly to the mortality rate of seasonal influenza. The lower panel shows the cumulative deaths over all States, in terms of the posterior expectation (blue line) and confidence intervals (shaded area). The (black dots) correspond to empirical data. The confidence intervals in these figures should not be overinterpreted: they were approximated (under large-number assumptions) by a Poisson distribution. This approximation was used purely

for computational expediency (and was not used during model inversion).

The underlying causes of the (predicted) outcomes in Figure 8 are shown in Figure 9. The predicted outcomes are reproduced in the upper panels, in terms of rates per day (upper left panel) and cumulative cases and deaths (upper right panel). In addition, the model has generated the occupancy of critical care unit beds that accompanies these State-specific predictions. The lines correspond to the predicted numbers under a model with a regional social distancing strategy. The dots correspond to the data observed so far. These trajectories are generated by fluctuations in the probability of being away from home (i.e., denoted by the state *work*, **Lw**), whether one is infectious or not, the clinical expression of the infection and the probability that one tests negative or positive. A key aspect of these results is the rapid acquisition of herd immunity to levels of about 30% at the peak of the first wave (see yellow lines in the middle right panel of Figure 9). This rise is most pronounced during the early phase of the pandemic that shows subsequent fluctuations due to movement between states and a mild loss of immunity.

One can see two sorts of second wave, in terms of new cases and deaths in the upper left panel. The first is an early secondary peak about four weeks after the first peak (orange arrow). This is particularly evident in the new cases predicted for New York. This coincides with a relaxation of social distancing and a concomitant influx of people from other States (compare this with the second wave of new cases in Figure 1). However, there is a more protracted and pronounced second wave after 28 weeks (blue arrow) that induces a fluctuation in social distancing behaviour; again, most evident in New York. The mechanism for this second peak is due to the endogenous loss of immunity. Increasing the period of immunity from 32 months to 8 years delays this peak by about 20 weeks (data not shown). As we will see next, decreasing the period of immunity to 4 months accelerates the second peak so that it encroaches on the first.

Finally, note that social distancing never disappears, i.e., the blue lines reporting the probability of leaving home (left middle panel) never quite return to their pre-pandemic levels. In other words, there is a persistent failure to return to levels of work prior to the pandemic. This persistent change in social distancing behaviour is an integral part of the endemic equilibrium simulated in these analyses. Put simply, there is an equilibrium that constitutes the endpoint of any 'exit strategy'—and this equilibrium will be attained at about 40 weeks (i.e., 10 months) following the onset of the pandemic. However, this will not be a return to normal life; it will be a way of 'living with COVID-19'. This way of living is comfortably within the resources of our ability to provide critical care for those people who need it. Furthermore, as noted above, this endemic equilibrium entails mortality rates due to COVID-19 that are less than half those due to seasonal influenza. It should be reiterated that these are just model predictions and should not be taken literally. They should be read as the kind of predictions that can be made with suitable modelling[14].

A key aspect of the simulation results is that new cases in the second wave are not necessarily accompanied by marked increases in daily death rates. Furthermore, the amplitude of the second wave, under this model, is much less than that of the first. This reflects the immunity acquired during the first wave. In other words, the context in which the second wave emerges is contextualised by the immunological memory of the first exposure to the virus. In these simulations, herd immunity is between 40% and 70%. These simulations can be compared with the predictions of seasonal re-emergencies of COVID-19, based upon models that factor in seasonal variations in transmission, of the kind associated with seasonal influenza. For example, short term immunity on the order of 10 months (similar to HCoV-OC43 and HCoV-HKU1) would engender annual SARS-CoV-2 outbreaks, while longer-term immunity (24 months) favours biennial outbreaks (Kissler et al., 2020).

---

[14] Note also that we have not included any therapeutic advances or vaccination programs in this model.

The simulations above should be compared with the corresponding simulations in Figure 10 under short term immunity. Here, we repeated the analysis but changed the period of immunity from 32 months to 4 months. In this scenario, herd immunity declines rapidly, and endemic equilibrium is reached almost as soon as the first wave subsides. This endemic equilibrium is probably not sustainable: although the demand for critical care does not exceed total capacity, this demand is mitigated by persistent social distancing. In turn, this social distancing precludes any meaningful return to work (in the majority of States simulated, only 8% or less of the population returns to work). On the assumption that the ensuing damage to social and economic infrastructure is untenable, the scenario of losing immunity within months is difficult to countenance.

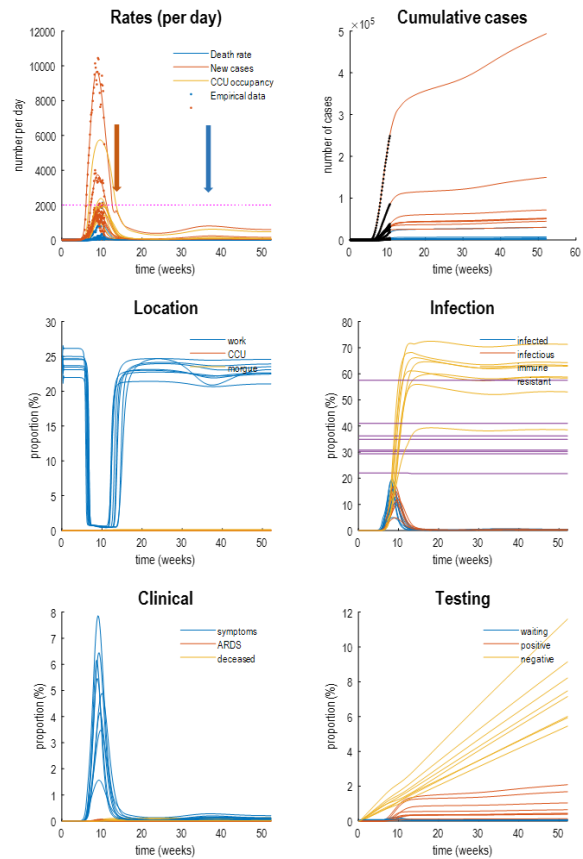

FIGURE 9

**Figure 9: the pandemic under long-term immunity (32 months)**. This figure reports the predicted outcomes and underlying latent states generating those outcomes over a one-year period, under a regional social distancing strategy. The upper panels show some key outcomes, some of which are measurable. Here, daily death rates are shown in blue, new cases in red and CCU occupancy in orange. The lines correspond to the predictions of the model, while the dots are empirical data available at the time of writing (18th of April 2020). The dotted line in the upper left panel corresponds to the typical critical care capacity of a large city. The same results are shown on the upper right panel in terms of cumulative new cases (red) and deaths (blue). The underlying or latent causes of this mortality are shown in the lower panels. These are organised according to the four factors of the generative model. In each panel, the latent states are plotted for the eight States considered in this analysis. The location factor shows that, under this strategy, the number of people away from the home (or equivalent location) decreases sharply at the onset of the outbreak and then recovers slowly over the ensuing weeks. In terms of infection, there is a rapid acquisition of immunity, to varying levels between 40% and 80% over the first months of the pandemic. At any one time, about 18% or less of the population is either infected or infectious. In terms of the clinical expression of these infections, 8% percent or less of people will experience symptoms and a small minority will progress to acute respiratory distress, from which they may recover or die. Under this model, positive test results for the virus (based on buccal swabs) accumulate over time as more and more people are tested. In the initial phases of the outbreak, most people are negative. However, during the onset of the pandemic about a half to a third of people tested are positive. This proportion declines over the ensuing months.

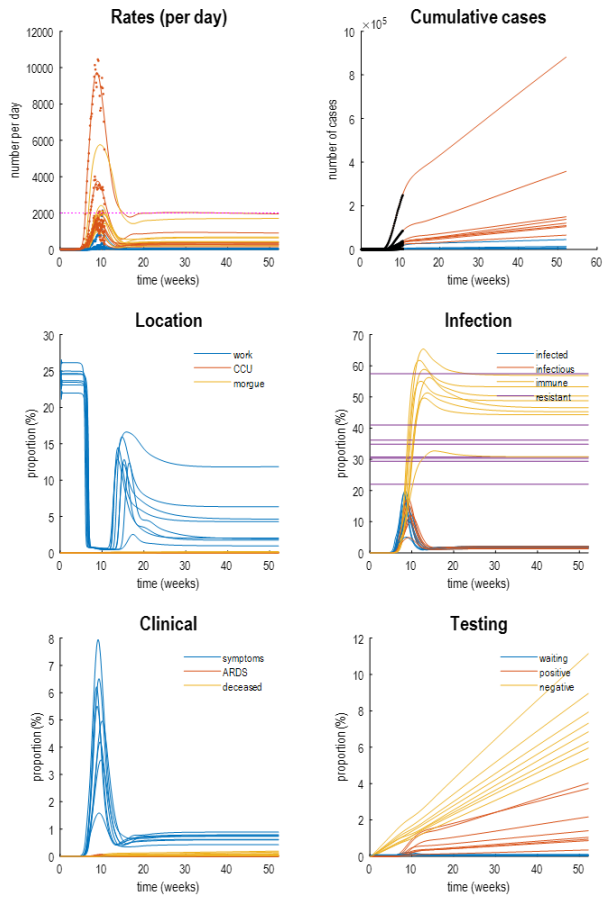

FIGURE 10

**Figure 10**Figure 9**: the pandemic under short-term immunity (4 months)**. This figure uses the same format as the previous figure. The only difference here is that we decreased the period of immunity from 32 months to 4 months.

The emerging picture is that two things are crucially important in shaping the long-term trajectory of the pandemic. First, the acquisition of herd immunity and second, the rate at which immunity is lost. At present, the second factor is difficult to assess empirically, which is why cases of reinfection would be so telling[15]. Documented cases of reinfection speak to a short-term immunity. At the time of writing, there is little evidence to suggest reinfection is a characteristic of COVID-19. Indeed, the evidence points in the other direction (Bao et al., 2020). So, is there any evidence for a rapid acquisition of herd immunity? At the time of writing, there are no published reports; however, preprints and local media have identified apposite studies in California. These studies are in a position to provide important data that will endorse or constrain the modelling of immunity. Figure 11 shows the predicted immunity for people in California shown in

---

[15]  Note that that a confirmed reinfection is distinct from a second positive test, which might be expected given the false positive (and negative) rates of many tests.

Figure 9. These predictions are not inconsistent with early reports of antibody testing in California based upon preprints (Bendavid et al., 2020)[16] and local media reports[17], shown as confidence intervals (vertical bars) and point estimates (red dot) respectively. Although the consilience should not be overinterpreted, these provisional findings are in line with the model predictions; although the Santa Clara prevalence of antibodies is about half what would have been predicted on the basis of the current model. As noted in the LA Times, these results suggest:

"*[t]he fatality rate may be much lower than previously thought. But although the virus may be more widespread, the infection rate still falls far short of herd immunity that, absent a vaccine, would be key to return to normal life*." ([https://www.latimes.com/california/story/2020-04-20/coronavirus-serology-testing-la-county](https://www.latimes.com/california/story/2020-04-20/coronavirus-serology-testing-la-county))

In fact, on the current reading of the model predictions, this prevalence of seropositive cases in California is consistent with the acquisition of herd immunity over the next few weeks. Clearly, it would be reassuring—if not imperative—to have more definitive data from other States (or countries) to track herd immunity as it develops.

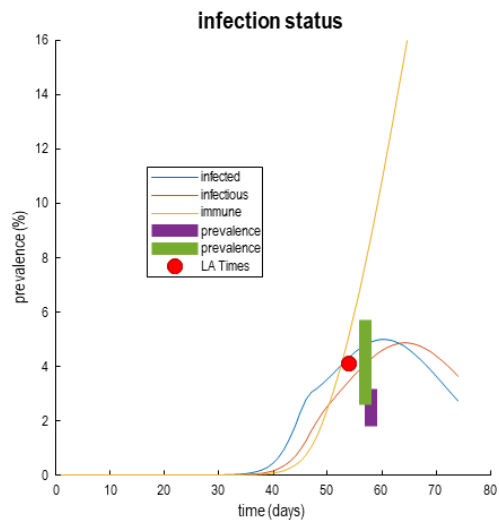

FIGURE 11

---

[16]  The authors measured the seroprevalence of antibodies to SARS-CoV-2 in Santa Clara County on the third and fourth of April 2020, using a lateral flow immunoassay. Under three scenarios for their test performance, the population prevalence of COVID-19 in Santa Clara ranged from 2.49% (95CI 1.80-3.17%) to 4.16% (2.58-5.70%).

[17]  Initial results from the first large-scale study tracking the spread of the coronavirus in LA county found that 4.1% of adults have antibodies. This translates to roughly 221,000 to 442,000 adults who have recovered from an infection. LA county had reported fewer than 8,000 cases at that time. Our thanks to Virginia Webber for forwarding this material.

**Figure 11: predicted and observed levels of immunity**. This figure reproduces the results of Figure 9, with a focus on the states of the *infection* factor. The blue line reports the expected percentage of the population who are infected in California, up until the date of writing. The red line shows the prevalence of contagious or infectious people that follows a few days later. The yellow line reports the acquisition of immunity; namely, the number of people who have moved from a *contagious* state to an *immune* state. The two bars represent the 90% confidence intervals from an early report from Santa Clara. The red dot corresponds to the estimated prevalence of immunity according to a local media report. Please see main text for details.

In summary, the acquiring and maintaining critical levels of immunity are crucial factors in determining the course of the pandemic over the next few months. Put simply, the rate at which we move to (a potentially catastrophic) endemic equilibrium will be much faster if immunity is lost quickly (via antigenic drift or mixing of the immune pool)— sufficiently quickly to preclude a second wave. This concludes our summary of the results under models of regional social distancing. In the next section, we turn to other aspects of strategic responses and ask what are the best mitigation strategies, under the current model?

## Mitigation strategies

In this section, we use the parameter estimates from the DCM to integrate 18 months into the future and record various outcomes under distinct social distancing strategies. The parameters of these strategies are estimated from empirical data; however, their functional form is a question of model selection. Having said this, given that there is only data from the initial phase of the pandemic, model comparison may be best left until after the epidemic has run its course. In the interim, we can examine the effect of strategic responses on metrics of interest; here, the total number of deaths, the total number of working days and the peak occupancy of critical care facilities. We will consider mitigation strategies that are driven by the prevalence of infection in any given State or whether national (i.e., Federal) measures of prevalence inform our behaviour. First, we look at the effect of relaxing social distancing (by increasing the social distancing threshold) and then repeat the analysis under a Federal policy.

Figure 12 shows the results of simulations under 16 levels of social distancing based upon regional prevalence. The different levels were simulated by scaling the social distancing threshold of about 3% in Figure 5 from very small values (around 0.05%) to very large values (around 100%). Figure 12 uses the same format as previous figures to quantify the rate of new cases and deaths per day (upper left panel), cumulated cases (upper right panel) and the underlying or latent causes (lower four panels). Here, the lines report the average rates and probabilities over States, for different levels of social distancing. With stringent social distancing (i.e., a low threshold) the initial relaxing of the lockdown after the first wave is quickly reversed as new cases start to accumulate. The probability of returning to work peaks at about 20% and then falls to negligible levels as the months pass. In this scenario, the second wave is delayed until about 60 weeks following the first wave. As social distancing is relaxed by increasing the threshold to its posterior estimate from the original model estimation, the return to work approaches endemic equilibria at successively higher levels, until, at low levels of social distancing (high thresholds) there is hardly any fluctuation in the probability of being found at work. At the same time, a second peak at around 40 weeks (30 weeks after the first peak) emerges and cumulative cases and deaths rise more quickly over the months. This spectrum of long-term trajectories is summarised by the green and blue arrows illustrating the suppression of a second wave at 70 weeks and the emergence of an earlier wave at 40 weeks.

The key aspect of these simulations is that there is a balance between an overzealous lockdown that precludes any meaningful return to work and a capricious social distancing strategy that releases a second wave a few months after the first. This effect of social distancing is illustrated more explicitly in Figure 13

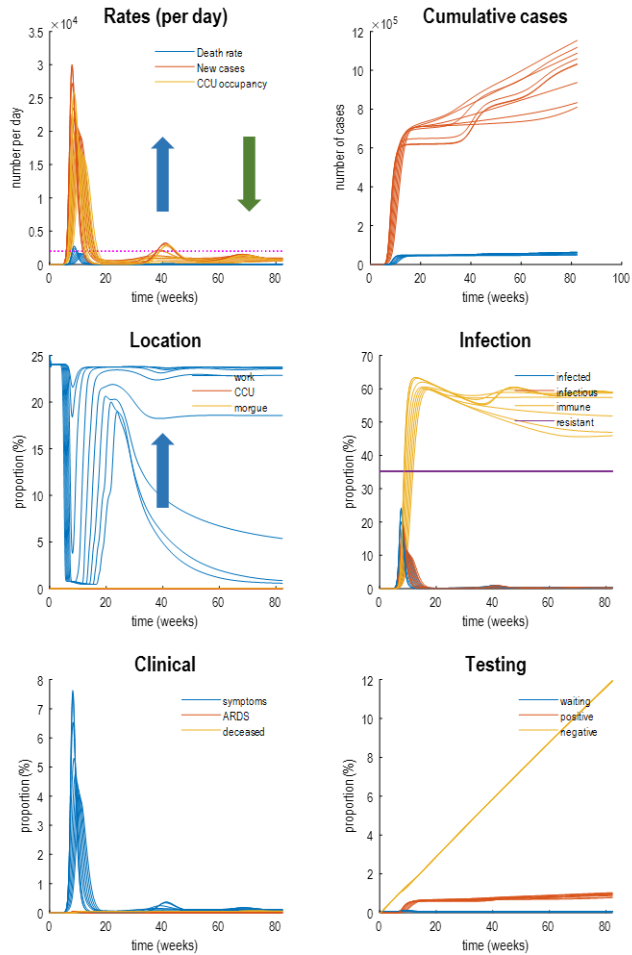

**Figure 12: the effect of social distancing**. This figure shows the results of simulations under different levels of social distancing, based upon regional prevalence. This figure uses the same format as previous figures, to illustrate the rate of new cases and deaths per day (upper left panel), accumulated cases (upper right panel) and the underlying or latent causes (lower four panels). Here, the lines report the average rates and probabilities over States, for different levels of social distancing. These levels were evaluated in 16 steps by scaling the posterior expectation of the social distancing thresholds from exp(-4) to exp(4) = 54.6. The key effects of this scaling are summarised in the next figure.

Figure 13 plots the cumulative deaths, working days and lost weeks as a function of the (logarithmic) deviation from an estimated social distancing threshold of about 3%. As one might intuit, increasing the threshold monotonically increases the number of working days at the expense of cumulative deaths, ranging across the eight states from about 48,000 to 62,000, over a period of 18 months. Changing the threshold for social distancing has a nonlinear aspect. This can be evinced more clearly by plotting the weeks lost due to lockdown as a function of the (logarithmic) changes in threshold. Weeks lost was quantified in terms of the number of weeks, over 18 months, during which the probability of being at work was less than 8%. The lower left panel of Figure 13 highlights the switch from lockdown strategies—which preclude a long-term return to work—to those that permit near-normal social distancing in the long term. Interestingly, the empirical (posterior expectation) estimate of social distancing based upon the data is characterised by a threshold that is close to the transition between the two kind of strategies. Again, as one might intuit, these simulations speak to a trade-off between lives and weeks lost due to the pandemic—and our adaptive response (see lower right panel).

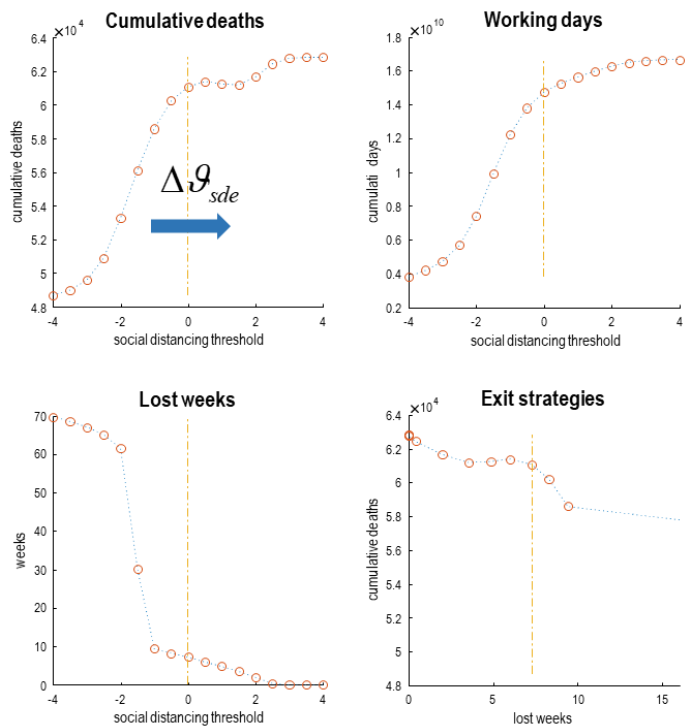

**Figure 13**

**Figure 13: lost lives and weeks**. This figure summarises the key effects of social distancing on cumulative deaths and working days lost due to the virus. The upper panels plot the cumulative deaths and days at work under different levels of social distancing. As in the previous figure, social distancing was evaluated over 16 levels by scaling the threshold estimated for each State. In terms of log parameters, this corresponds to adding or subtracting a change (between -4 and +4 natural units). In this figure, social distancing is expressed in terms of these changes, where zero

corresponds to no change from the (posterior) estimates of the previous section. The lower panel provides another perspective on working days lost by calculating the number of lost weeks—defined operationally as the number of weeks during which the probability of going to work was less than 8%. The final (lower right) panel plots cumulative deaths against weeks lost. This illustrates the trade-off between the loss of life and working weeks—reflected in the decreasing monotonic relationship between these two outcomes.

Finally, we repeated the above analysis under regional and national strategies, using high (1/4) and low (1/32) social distancing thresholds. Figure 14 (upper row) shows the effects of the four kinds of strategy on cumulative deaths, total number of working days and peak CCU occupancy. It is apparent that a hard (low threshold) strategy reduces deaths and frees up working days to a greater extent than soft (high threshold) strategies. Interestingly, a regional response strategy based on local prevalence rates incurs fewer deaths with a slight increase in CCU occupancy. There is no discernible difference between a regional and national strategy on working days.

The middle row shows the same results but when immunity is lost over four months, as opposed to 32 months. As one might intuit, increasing the rate at which immunity is lost substantially increases death rates and other costs. In this scenario, the impact of a hard (low threshold) lockdown strategy on cumulative deaths is more marked. And the relative benefit of a regional versus national policy is more pronounced. This is at the expense of damage to the economy, in terms of the relative number of days lost. Total CCU occupancy is largely unaffected with short-term immunity (because the loss of immunity only affects the epidemiology after it is acquired during the first peak).

The lower row reproduces the upper row, with the inclusion of a back-to-work policy, in which people who were seropositive are allowed to leave the home. The point made by these simulations is that a back-to-work policy has limited effects on mortality rates[18] or peak CCU occupancy but greatly ameliorates the economic damage entailed by the lost working days.

---

[18] Although it appears to reverse the effect of a regional versus national policy.

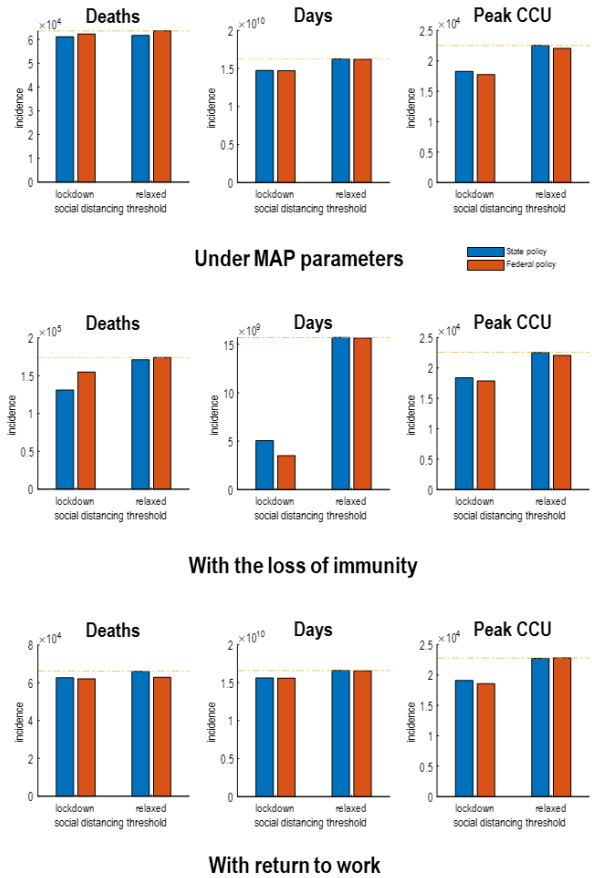

**Figure 14: different strategies evaluated**. These bar charts report the effects of different strategies on cumulative deaths (left column), total number of working days (middle column) and peak occupancy of CCU (right column). The top row shows simulation results based upon the parameters used in the previous figures. One can see that a low threshold (lockdown) strategy generally reduces cumulative deaths and working days lost to the economy, while augmenting peak CCU occupancy. The middle row reproduces the same analysis but under more pessimistic assumptions about the retention of immunity. Specifically, we reduced the time constant for retaining immunity from 32 months to 4 months. Under this scenario, death rates increase dramatically (approximately doubling), with an accompanying loss of working days. Peak CCU is largely unaffected, because this parameter determines long-term outcomes or trajectories as opposed to initial responses. The lower row shows the equivalent results when including a back-to-work policy based upon serological testing. The only effect of this, under the current model, is to exacerbate the effect of a hard versus relaxed social distancing strategy on the number of working days lost. This follows because returning people who are immune to the community has no effect on morbidity or for critical care; however, it does take the pressure off the economy.

# Conclusion

This report describes an extension of a single region (epidemic) model that furnishes a (pandemic) model of regions that collectively participate in a pandemic. Our focus has been on the genesis of a second wave of new cases—and potential deaths—due to the loss of immunity within a regional population and the influx of people from other regions. We have showcased this model by applying it to statistics from the United States, treating each State as a separable region. When using the optimised model parameters one can simulate the impact of various strategies or (non-pharmaceutical) responses.

If one subscribes to the modelling in this report, then there are several narratives one could entertain. These depend sensitively on the rate at which the immunity is lost. If SARS-CoV-2 confers immunity that lasts for years, the following narrative might be appropriate:

*"Endemic equilibrium will be reached by the end of the year, by which time COVID-19 will become 'another way to die'. COVID-19 will account for a small proportion of deaths—with mortality rates that are less than seasonal influenza (approximately 1/3). This state of affairs rests upon an adaptive (i.e., reactive) response to intervening fluctuations in the prevalence of infection (and demands upon critical care). Operationally, this response could be characterised by social distancing when either the prevalence of infection in the population exceeds 3% for the number of COVID-19 patients in critical care surpasses a quarter of maximum capacity. Comprehensive social distancing (i.e. lockdown) is currently in force and will last, on average, about 7.5 weeks in each affected region. For most countries, at the time of writing, this suggests a relaxing of lockdown in 3 weeks.*

*However, this relaxation will not return to pre-pandemic levels of interpersonal contact. In other words, there will be an enduring pressure to reduce interpersonal contact, which will reduce the time spent in the company of others by 5% or less. The road to equilibrium will be relatively smooth with a slight bump (second wave) at about seven months following the initial outbreak (i.e., November). This will not require lockdown but there will be an appreciable increase in the number of cases and critical care uptake. This second wave should last for about five weeks and may be confounded by the onset of a flu season."*

Under this narrative, social distancing will become 'a way of living' with COVID-19 and reflect changing attitudes to prosocial behaviour; very much like our attitudes to recycling, smoking or outdated and dangerous practices, such as the use of laudanum or mercury in the Victorian age. For example, unnecessary commuting and international meetings may be seen as antisocial and unhygienic. Similarly, shaking hands may become as socially sanctioned as frotteurism. In short, changing attitudes, affordances and dispositions may be sufficient social distancing mechanisms to guarantee an endemic equilibrium. Note that the above narrative makes no mention of 'exit strategies', vaccination, or anti-viral therapy. This is because strategic responses are modelled as an inherent part of the epidemiology. In other words, they are treated as part of the process as a *reactive* strategy, as opposed to a *proactive* strategy. This does not mean that governments are under no pressure to declare their exit strategies. Rather, this declaration is part of the process of changing attitudes to social behaviour.

As in (Friston et al., 2020), this narrative is not a prediction. It is a concrete example of the kind of prediction on offer, with a suitably formulated and informed epidemiological model that installs social behaviour into the dynamics. This particular narrative depends on long-lasting immunity. A different story would be told if immunity to SARS-CoV-2 is lost within months. This speaks to two pressing issues. First, has the first wave induced a substantive herd immunity predicted by the DCM? Second, if this immunity has been induced, how long will it last? The answer to the first question should be available within the next few weeks, as studies assessing community levels of seroconversion appear. The answer to the second question is more vexed because—from a purely epidemiological perspective—one might have to wait and see (given the many factors that determine the loss of immunity and potential variation within different cohorts). In the final report of this series, we will estimate the period of immunity by comparing models over a range of short-term immunity, in terms of their model evidence. At some point over the next month or so, there should be sufficient data to render this model comparison sufficiently definitive to estimate the effective period of immunity one can expect.

As with all dynamic causal modelling studies, everything is entirely conditioned upon the models that have been evaluated in terms of their evidence. In this report, the aim of the modelling is not to provide predictions or guidance *per se*—it is to show that such predictions are possible under a suitably configured model using state-of-the-art variational Bayesian model inversion and reduction. In other words, one can reproduce these analyses under different models or prior assumptions in a few minutes on a personal computer. This allows one to explore different models in an efficient fashion; thereby treating the modelling as hypotheses testing, as more data becomes available. Being able to identify the best hypothesis or model—in terms of its parameterisation of structure—is potentially important. This is because the model and (maximum *a posteriori*) parameters can then form the basis of a prediction about data that has not yet been observed, i.e., the future. In this light, we will not list the shortcomings of this particular model. Any 'shortcoming' is just a statement of an alternative model that can be included in the model comparison procedure to optimise the model in and of itself.

Perhaps the most important aspect of this modelling is its focus on herd immunity and the notion of an adaptive or reactive strategy. As in our provisional modelling of a single outbreak in a single region, the passage from the onset of an outbreak to endemic equilibrium depends on establishing herd immunity to a greater or lesser extent. This underwrites the repeated call for a greater understanding of the virology and immunology of COVID-19. Not only will this speak to therapeutic interventions, but simply being able to measure the proportion of individuals in a population who are immune would provide informative constraints on modelling—and subsequent projections. The second aspect—of adaptive social distancing—can be contrasted with other formulations that prescribe a fixed pattern of social distancing. For example, a sawtooth or intermittent social distancing that enables the acquisition of herd immunity without overwhelming critical care capacity (Ferguson et al., 2006). Both adaptive and proactive schedules (e.g., encouraging the use of facemasks) have their merits and could be evaluated using the procedures outlined above.

Adaptive strategies provide a quantitative and formal guideline for personal and governmental responses as the pandemic develops. In other words, the response becomes part of the ensemble dynamics that determine the eventual outcome. A pragmatic advantage of adaptive strategies (see Figure 4) is that they can be operationalised given available estimates of the prevalence. For a threshold strategy, one can simply revert to one mode of social distancing or another, whenever a particular threshold is passed. An interesting perspective on adaptive strategies follows from the fact that—in the model—they are an integral part of the process. In other fields, this corresponds to the notion of Chaos control; e.g., (Rose, 2014) and may yield to an optimal control theoretic treatment (Fleming and Sheu, 2002; Kappen, 2005; Todorov and Jordan, 2002). In other words, the optimal policy can be treated very much like an engineering problem or, indeed, an active inference problem. This brings us to our last point.

In closing, there is one perspective on applications of generative models to pandemics that touches on (unrelated) work in theoretical neurobiology. This work tries to understand the behaviour of sentient systems, such as ourselves, in terms of active inference and the optimisation of variational free energy associated with our internal world models (Parr and Friston, 2018). On this view, using a generative model of a pandemic to inform policies becomes formally identical to the imperatives that underwrite active inference (Clark, 2016; Hohwy, 2013). In brief, these imperatives are to bring about those situations that we find the least surprising; namely, that we all elude death and return to normal.

## Software note

The figures in this report can be reproduced using annotated (MATLAB/Octave) code that is available as part of the free and open source academic software SPM (https://www.fil.ion.ucl.ac.uk/spm/), released under the terms of the GNU General Public License version 2 or later. The routines are called by a demonstration script that can be invoked by typing >> DEM_COVID_X at the MATLAB prompt. At the time of writing, these routines are undergoing software validation in our internal source version control system—that will be released in the next public release of SPM (and via GitHub at https://github.com/spm/). In the interim, please see https://www.fil.ion.ucl.ac.uk/spm/COVID-19/.

The data used in this technical report are available for academic research purposes from the 2019 Novel Coronavirus COVID-19 (2019-nCoV) Data Repository by Johns Hopkins CSSE, hosted on GitHub at https://github.com/CSSEGISandData/COVID-19. The time series tables from States in America used in this report are stored in files time_series_COVID19_confirmed_US.csv and time_series_COVID19_deaths_US.csv.

## Acknowledgements

This work was undertaken by members of the Wellcome Centre for Human Neuroimaging, UCL Queen Square Institute of Neurology. The Wellcome Centre for Human Neuroimaging is supported by core funding from Wellcome [203147/Z/16/Z]. A.R. is funded by the Australian Research Council (Refs: DE170100128 and DP200100757). A.J.B. is supported by a Wellcome Trust grant WT091681MA. CL is supported by an MRC Clinician Scientist award (MR/R006504/1).

The authors declare no conflicts of interest.

# References

Bao, L., Deng, W., Gao, H., Xiao, C., Liu, J., Xue, J., Lv, Q., Liu, J., Yu, P., Xu, Y., Qi, F., Qu, Y., Li, F., Xiang, Z., Yu, H., Gong, S., Liu, M., Wang, G., Wang, S., Song, Z., Zhao, W., Han, Y., Zhao, L., Liu, X., Wei, Q., Qin, C., 2020. Reinfection could not occur in SARS-CoV-2 infected rhesus macaques. bioRxiv, 2020.2003.2013.990226.

Bendavid, E., Mulaney, B., Sood, N., Shah, S., Ling, E., Bromley-Dulfano, R., Lai, C., Weissberg, Z., Saavedra, R., Tedrow, J., Tversky, D., Bogan, A., Kupiec, T., Eichner, D., Gupta, R., Ioannidis, J., Bhattacharya, J., 2020. COVID-19 Antibody Seroprevalence in Santa Clara County, California. medRxiv, 2020.2004.20062463.

Beutler, B., Eidenschenk, C., Crozat, K., Imler, J.L., Takeuchi, O., Hoffmann, J.A., Akira, S., 2007. Genetic analysis of resistance to viral infection. Nature reviews. Immunology 7, 753-766.

Chan, K.H., Chan, J.F., Tse, H., Chen, H., Lau, C.C., Cai, J.P., Tsang, A.K., Xiao, X., To, K.K., Lau, S.K., Woo, P.C., Zheng, B.J., Wang, M., Yuen, K.Y., 2013. Cross-reactive antibodies in convalescent SARS patients' sera against the emerging novel human coronavirus EMC (2012) by both immunofluorescent and neutralizing antibody tests. The Journal of infection 67, 130-140.

Chinazzi, M., Davis, J.T., Ajelli, M., Gioannini, C., Litvinova, M., Merler, S., Pastore y Piontti, A., Mu, K., Rossi, L., Sun, K., Viboud, C., Xiong, X., Yu, H., Halloran, M.E., Longini, I.M., Vespignani, A., 2020. The effect of travel restrictions on the spread of the 2019 novel coronavirus (COVID-19) outbreak. Science, eaba9757.

Clark, A., 2016. Surfing Uncertainty: Prediction, Action, and the Embodied Mind. Oxford University Press.

Cristiani, L., Mancino, E., Matera, L., Nenna, R., Pierangeli, A., Scagnolari, C., Midulla, F., 2020. Will children reveal their secret? The coronavirus dilemma. European Respiratory Journal, 2000749.

Ferguson, N., Laydon, D., Nedjati Gilani, G., Imai, N., Ainslie, K., Baguelin, M., Bhatia, S., Boonyasiri, A., Cucunuba Perez, Z., Cuomo-Dannenburg, G., 2020. Report 9: Impact of non-pharmaceutical interventions (NPIs) to reduce COVID19 mortality and healthcare demand.

Ferguson, N.M., Cummings, D.A., Fraser, C., Cajka, J.C., Cooley, P.C., Burke, D.S., 2006. Strategies for mitigating an influenza pandemic. Nature 442, 448-452.

Fleming, W.H., Sheu, S.J., 2002. Risk-sensitive control and an optimal investment model II. Ann. Appl. Probab. 12, 730-767.

Friston, K., Parr, T., Zeidman, P., 2018. Bayesian model reduction. arXiv preprint arXiv:1805.07092.

Friston, K., Zeidman, P., Litvak, V., 2015. Empirical Bayes for DCM: A Group Inversion Scheme. Frontiers in systems neuroscience 9, 164.

Friston, K.J., Parr, T., Zeidman, P., Razi, A., Flandin, G., Daunizeau, J., Hulme, O.J., Billig, A.J., Litvak, V., Moran, R.J., Price, C.J., Lambert, C., 2020. Dynamic causal modelling of COVID-19. arXiv e-prints, arXiv:2004.04463.

Gibson, J.J., 1977. The theory of affordances, in: R, S., Bransford, J. (Eds.), Perceiving, acting, and knowing: Toward an ecological psychology. Erlbaum, Hillsdale, NJ, pp. 67-82.

Heer, C.D., Sanderson, D.J., Alhammad, Y.M.O., Schmidt, M.S., Trammell, S.A.J., Perlman, S., Cohen, M.S., Fehr, A.R., Brenner, C., 2020. Coronavirus Infection and PARP Expression Dysregulate the NAD Metabolome: A Potentially Actionable Component of Innate Immunity. bioRxiv, 2020.2004.2017.047480.

Hohwy, J., 2013. The Predictive Mind. Oxford University Press, Oxford.

Huang, C.L., Wang, Y.M., Li, X.W., Ren, L.L., Zhao, J.P., Hu, Y., Zhang, L., Fan, G.H., Xu, J.Y., Gu, X.Y., Cheng, Z.S., Yu, T., Xia, J.A., Wei, Y., Wu, W.J., Xie, X.L., Yin, W., Li, H., Liu, M., Xiao, Y., Gao, H., Guo, L., Xie, J.G., Wang, G.F., Jiang, R.M., Gao, Z.C., Jin, Q., Wang, J.W., Cao, B., 2020. Clinical features of patients infected with 2019 novel coronavirus in Wuhan, China. Lancet (London, England) 395, 497-506.

Jafri, H.H., Singh, R.K.B., Ramaswamy, R., 2016. Generalized synchrony of coupled stochastic processes with multiplicative noise. Physical Review E 94, 052216.

Kaluza, P., Meyer-Ortmanns, H., 2010. On the role of frustration in excitable systems. Chaos 20, 043111.

Kappen, H.J., 2005. Path integrals and symmetry breaking for optimal control theory. Journal of Statistical Mechanics: Theory and Experiment 11, P11011.

Kissler, S.M., Tedijanto, C., Goldstein, E., Grad, Y.H., Lipsitch, M., 2020. Projecting the transmission dynamics of SARS-CoV-2 through the postpandemic period. Science, eabb5793.

Ladenbauer, J., Obermayer, K., 2019. Weak electric fields promote resonance in neuronal spiking activity: Analytical results from two-compartment cell and network models. PLoS Comput Biol 15, e1006974.

Li, R., Pei, S., Chen, B., Song, Y., Zhang, T., Yang, W., Shaman, J., 2020. Substantial undocumented infection facilitates the rapid dissemination of novel coronavirus (SARS-CoV2). Science, eabb3221.

Lizarazu, M., Lallier, M., Molinaro, N., 2019. Phase-amplitude coupling between theta and gamma oscillations adapts to speech rate. Ann N Y Acad Sci.

Mizumoto, K., Chowell, G., 2020. Estimating Risk for Death from 2019 Novel Coronavirus Disease, China, January-February 2020. Emerging infectious diseases 26.

Moghadas, S.M., Shoukat, A., Fitzpatrick, M.C., Wells, C.R., Sah, P., Pandey, A., Sachs, J.D., Wang, Z., Meyers, L.A., Singer, B.H., 2020. Projecting hospital utilization during the COVID-19 outbreaks in the United States. Proceedings of the National Academy of Sciences.

Moran, R.J., Fagerholm, E.D., Cullen, M., Daunizeau, J., Richardson, M.P., Williams, S., Turkheimer, F., Leech, R., Friston, K.J., 2020. Estimating required 'lockdown' cycles before immunity to SARS-CoV-2: Model-based analyses of susceptible population sizes, 'S0', in seven European countries including the UK and Ireland. arXiv e-prints, arXiv:2004.05060.

Nishiura, H., Linton, N.M., Akhmetzhanov, A.R., 2020. Serial interval of novel coronavirus (COVID-19) infections. International journal of infectious diseases.

Paget, J., Spreeuwenberg, P., Charu, V., Taylor, R.J., Iuliano, A.D., Bresee, J., Simonsen, L., Viboud, C., Global Seasonal Influenza-associated Mortality Collaborator, N., Teams*, G.L.C., 2019. Global mortality associated with seasonal influenza epidemics: New burden estimates and predictors from the GLaMOR Project. J Glob Health 9, 020421-020421.

Parr, T., Friston, K.J., 2018. The Anatomy of Inference: Generative Models and Brain Structure. Frontiers in computational neuroscience 12.

Prem, K., Liu, Y., Russell, T.W., Kucharski, A.J., Eggo, R.M., Davies, N., Flasche, S., Clifford, S., Pearson, C.A., Munday, J.D., 2020. The effect of control strategies to reduce social mixing on outcomes of the COVID-19 epidemic in Wuhan, China: a modelling study. The Lancet Public Health.

Rose, N.R., 2014. Learning from myocarditis: mimicry, chaos and black holes. F1000prime reports 6, 25.

Russell, T.W., Hellewell, J., Jarvis, C.I., van Zandvoort, K., Abbott, S., Ratnayake, R., Cmmid Covid-Working, G., Flasche, S., Eggo, R.M., Edmunds, W.J., Kucharski, A.J., 2020. Estimating the infection and case fatality ratio for coronavirus disease (COVID-19) using age-adjusted data from the outbreak on the Diamond Princess cruise ship, February 2020. Euro surveillance : bulletin Europeen sur les maladies transmissibles = European communicable disease bulletin 25.

Schumacher, J., Wunderle, T., Fries, P., Jakel, F., Pipa, G., 2015. A Statistical Framework to Infer Delay and Direction of Information Flow from Measurements of Complex Systems. Neural Comput 27, 1555-1608.

Simonsen, L., Chowell, G., Andreasen, V., Gaffey, R., Barry, J., Olson, D., Viboud, C., 2018. A review of the 1918 herald pandemic wave: importance for contemporary pandemic response strategies. Annals of epidemiology 28, 281-288.

Steven, S., Yen Ting, L., Chonggang, X., Ethan, R.-S., Nick, H., Ruian, K., 2020. High Contagiousness and Rapid Spread of Severe Acute Respiratory Syndrome Coronavirus 2. Emerging Infectious Disease journal 26.

Su, S., Wong, G., Shi, W., Liu, J., Lai, A.C.K., Zhou, J., Liu, W., Bi, Y., Gao, G.F., 2016. Epidemiology, Genetic Recombination, and Pathogenesis of Coronaviruses. Trends in microbiology 24, 490-502.

Sun, K., Chen, J., Viboud, C., 2020. Early epidemiological analysis of the coronavirus disease 2019 outbreak based on crowdsourced data: a population-level observational study. The Lancet Digital Health.

Todorov, E., Jordan, M.I., 2002. Optimal feedback control as a theory of motor coordination. Nat Neurosci. 5, 1226-1235.

Verity, R., Okell, L.C., Dorigatti, I., Winskill, P., Whittaker, C., Imai, N., Cuomo-Dannenburg, G., Thompson, H., Walker, P.G.T., Fu, H., Dighe, A., Griffin, J.T., Baguelin, M., Bhatia, S., Boonyasiri, A., Cori, A., Cucunuba, Z., FitzJohn, R., Gaythorpe, K., Green, W., Hamlet, A., Hinsley, W., Laydon, D., Nedjati-Gilani, G., Riley, S., van Elsland, S., Volz, E., Wang, H., Wang, Y., Xi, X., Donnelly, C.A., Ghani, A.C., Ferguson, N.M., 2020. Estimates of the severity of coronavirus disease 2019: a model-based analysis. The Lancet. Infectious diseases.

Wang, C., Liu, L., Hao, X., Guo, H., Wang, Q., Huang, J., He, N., Yu, H., Lin, X., Pan, A., 2020a. Evolving Epidemiology and Impact of Non-pharmaceutical Interventions on the Outbreak of Coronavirus Disease 2019 in Wuhan, China. medRxiv.

Wang, D., Hu, B., Hu, C., Zhu, F., Liu, X., Zhang, J., Wang, B., Xiang, H., Cheng, Z., Xiong, Y., Zhao, Y., Li, Y., Wang, X., Peng, Z., 2020b. Clinical Characteristics of 138 Hospitalized Patients With 2019 Novel Coronavirus–Infected Pneumonia in Wuhan, China. JAMA 323, 1061-1069.

Wang, H., Wang, Z., Dong, Y., Chang, R., Xu, C., Yu, X., Zhang, S., Tsamlag, L., Shang, M., Huang, J., 2020c. Phase-adjusted estimation of the number of coronavirus disease 2019 cases in Wuhan, China. Cell Discovery 6, 1-8.

Wells, C.R., Sah, P., Moghadas, S.M., Pandey, A., Shoukat, A., Wang, Y., Wang, Z., Meyers, L.A., Singer, B.H., Galvani, A.P., 2020. Impact of international travel and border control measures on the global spread of the novel 2019 coronavirus outbreak. Proceedings of the National Academy of Sciences 117, 7504-7509.

Wölfel, R., Corman, V.M., Guggemos, W., Seilmaier, M., Zange, S., Müller, M.A., Niemeyer, D., Jones, T.C., Vollmar, P., Rothe, C., Hoelscher, M., Bleicker, T., Brünink, S., Schneider, J., Ehmann, R., Zwirglmaier, K., Drosten, C., Wendtner, C., 2020. Virological assessment of hospitalized patients with COVID-2019. Nature.

Wu, J.T., Leung, K., Leung, G.M., 2020. Nowcasting and forecasting the potential domestic and international spread of the 2019-nCoV outbreak originating in Wuhan, China: a modelling study. Lancet (London, England) 395, 689-697.

Yang, Z., Zeng, Z., Wang, K., Wong, S., Liang, W., Zanin, M., Liu, P., Cao, X., Gao, Z., Mai, Z., 2020. Modified SEIR and AI prediction of the epidemics trend of COVID-19 in China under public health interventions. Journal of Thoracic Disease 12.